\documentclass[journal]{IEEEtran}

\newcommand{\sqr}{\vrule height6pt width6pt depth1pt}
\def\qed{\hfill\sqr}

\usepackage{graphics} 
\usepackage{epsfig} 
\usepackage{times} 
\usepackage{amsmath} 
\usepackage{amssymb}  
\usepackage{setspace}
\usepackage{color}
\usepackage{fancyhdr}

\title{\LARGE \bf
Secret Key Generation for a Pairwise Independent Network Model}

\author{
         Sirin Nitinawarat, {\em Student Member, IEEE}, Chunxuan Ye, {\em Member,
         IEEE},\\
         Alexander Barg, Prakash Narayan, {\em Fellow, IEEE}, and
         Alex Reznik, {\em Member, IEEE}
\thanks{The work of S. Nitinawarat and P. Narayan was supported
by the National Science Foundation under Grants CCF0515124,
CCF0635271, CCF0830697, and InterDigital. The work of A. Barg
was supported by the National
Science Foundation under Grants CCF0515124, CCF0830699,
CCF0916919, DMS0807411, and InterDigital.  The
material in this paper was presented in parts at the IEEE
International Symposia on Information Theory, Nice, France, June
2007, and Toronto, Ontario, Canada, July 2008.}
\thanks{S. Nitinawarat, A. Barg and P. Narayan are with
the Department of Electrical and Computer Engineering and the
Institute for Systems Research, University of Maryland, College
Park, MD 20742, USA}
\thanks{ \hspace{-0.15in} Email: \{nitinawa, abarg, prakash\}@umd.edu.}
\thanks{C. Ye and A. Reznik are with InterDigital,
King of Prussia, PA 19406}
\thanks{ \hspace{-0.15in} Email:\{Chunxuan.Ye, Alex.Reznik\}@interdigital.com.}
}

\begin{document}

\maketitle

\thispagestyle{empty}
\renewcommand{\headrulewidth}{0.0pt}
\thispagestyle{fancy} \lhead{IEEE Transactions on Information
Theory, in revision prior to review for final approval.}

\begin{abstract}

We consider secret key generation for a ``pairwise independent
network'' model in which every pair of terminals observes
correlated sources that are independent of sources observed by all
other pairs of terminals. The terminals are then allowed to
communicate publicly with all such communication being observed by
all the terminals. The objective is to generate a secret key
shared by a given subset of terminals at the largest rate
possible, with the cooperation of any remaining terminals. Secrecy
is required from an eavesdropper that has access to the public
interterminal communication. A (single-letter) formula for secret
key capacity brings out a natural connection between the problem
of secret key generation and a combinatorial problem of maximal
packing of Steiner trees in an associated multigraph.  An explicit
algorithm is proposed for secret key generation based on a maximal
packing of Steiner trees in a multigraph; the corresponding
maximum rate of Steiner tree packing is thus a lower bound for the
secret key capacity.  When only two of the terminals or when all
the terminals seek to share a secret key, the mentioned algorithm
achieves secret key capacity in which case the bound is tight.


\end{abstract}

\section{Introduction}

Suppose that terminals $1, \ldots, m$ observe distinct but
correlated signals with the feature that every
pair of terminals observes a corresponding pair of correlated
signals that are independent of all other pairs of signals.
Following these observations, all the terminals can communicate
interactively over a public noiseless channel of unlimited capacity,
with all such communication being observed by all the terminals.  The
goal is to generate a secret key (SK), i.e., secret common
randomness, for a given subset $A$ of the terminals in
$\mathcal{M} = \{1, \ldots, m\}$ at the largest rate possible, with
secrecy being required from an eavesdropper that observes the
public interterminal communication.  All the terminals in
$\mathcal{M}$ cooperate in generating the SK for the secrecy-seeking
set $A$.

This model for SK generation, called a ``pairwise independent
network'' model, was introduced in \cite{YeRez07} (see also \cite{YeRezSha06}).
Abbreviated hereafter as the PIN model, it is motivated by practical
aspects of a wireless communication network in which terminals
communicate on the same frequency.  In a typical
multipath environment, the wireless channel between each pair of
terminals
produces a random mapping between the transmitted and received
signals which is time-varying and location-specific.  For a fixed
time and location, this mapping is reciprocal, i.e., effectively the
same in both directions.  Also, the mapping decorrelates over
different time-coherence intervals as well as over distances of the
order of a few wavelengths.

The PIN model is, in fact, a special case of a general multiterminal
``source model'' for secrecy generation studied by Csisz\'ar and
Narayan \cite{CsiNar04}.  The latter followed leading investigations
by Maurer \cite{Mau90, Mau93} and Ahlswede and Csisz\'ar
\cite{AhlCsi93} of SK generation by two terminals from
their correlated observations complemented by public communication.

A single-letter characterization of secret key capacity -- the largest
rate at which secrecy can be generated -- for the terminals in an
arbitrary subset $A$ of $\mathcal{M}$ was provided in \cite{CsiNar04}.
A particularization of this (general) SK capacity formula
to our PIN model displays the special feature that it
can be expressed in terms of a linear combination of mutual
information terms that involve only mutually independent pairs of
``reciprocal'' random variables (rvs).  Each such mutual information
term represents the maximum rate of an SK that can be generated
solely by a corresponding pair of terminals from only their own
observed signals using public communication \cite{Mau90, Mau93,
AhlCsi93}.  This observation leads to the following question that is
our main motivation:  {\em Can an SK of optimum rate for the
terminals in} $A$ {\em be generated by propagating mutually independent
and rate-optimal SKs for pairs of terminals in}
$\mathcal{M}$?

An examination of this question brings out points of contact
between SK generation for a PIN model and a combinatorial
problem of tree packing in a multigraph.  We propose an explicit
algorithm for propagating pairwise SKs for pairs of terminals in
$\mathcal{M}$ to form a groupwide SK for the terminals in
$A$.  This algorithm is based on a maximal packing of Steiner
trees (for $A$) in a multigraph associated with the PIN model.
Thus, the maximum rate of Steiner tree packing in this multigraph
is always a lower bound for SK capacity.  This bound is
shown to be tight when the secrecy-seeking set $A$ contains only
two terminals or when it consists of all the terminals.
In these situations, our
algorithm is capacity-achieving.  It is of independent interest to
note that given a
combinatorial problem of determining the maximum rate of Steiner
tree packing for $A$ in a multigraph, the SK capacity of
an associated PIN model provides, in reciprocity, an upper bound
for the mentioned rate, which is tight for the case
$|A|=2$ as well as for the spanning tree case $A = \mathcal{M}$.

In the study of secrecy generation for a multiterminal source
model, the notions of wiretap SK \cite{Mau90, Mau93,
AhlCsi93, CsiNar04} and private key \cite{CsiNar04} also have been
proposed.  The former notion corresponds to the eavesdropper having
additional access to a terminal not in the secrecy-seeking set $A$
and from which too the key must be concealed; this ``wiretapped''
terminal does not cooperate in secrecy generation.  A single-letter
characterization of the corresponding capacity remains unresolved
in general but for partial results and bounds (cf. e.g.,
\cite{AhlCsi93, Mau93, RenWo03, CsiNar04, GohAnan07, GohAnan08, CsiNar08}).
The notion of a private key is less restrictive, with the wiretapped
terminal being allowed to cooperate; the corresponding capacity is
known \cite{CsiNar04}.  We argue in Section IV
below that for a PIN model these two notions correspond to SK
generation for a reduced PIN model, thereby justifying our sole
focus on SK capacity.

Basic concepts and definitions are presented in Section II.
Section III contains statements of our results and proofs;
specifically, the SK capacity for the PIN model is given in
Section III.A, the connection of SK capacity with Steiner tree
packing is treated in Section III.B, and with spanning tree packing
in Section III.C.  Concluding remarks and pointers to a
sequel paper are contained in Section IV.

\section{Preliminaries}

We shall be concerned throughout with
a PIN model, which is a special case
of a general multiterminal
``source model'' for secrecy
generation with public communication
(see \cite{Mau93, AhlCsi93, CsiNar04, CsiNar08}).
Suppose that terminals $1, \ldots, m,\
m \geq 2,$ observe $n$ independent and identically distributed
(i.i.d.) repetitions of the rvs
$\tilde{X}_1, \ldots, \tilde{X}_m,$ denoted by $\tilde{X}_1^n,
\ldots, \tilde{X}_m^n,$ where $\tilde{X}_i^n = \left
(\tilde{X}_{i,1}, \ldots, \tilde{X}_{i,n} \right),\ i \in
\mathcal{M} = \{1, \ldots, m\}$. Each rv
$\tilde{X}_i,\ i \in \mathcal{M},$ is of the form $\tilde{X}_i
= \left( X_{ij},\ j \in \mathcal{M} \backslash \{i\} \right)$
with $m-1$ components, and the ``reciprocal pairs'' of rvs
$\{\left ( X_{ij}, X_{ji} \right ),\ 1 \leq i < j \leq m \}$
are mutually independent.  See Figure 1.
Thus, every pair of
terminals in $\mathcal{M}$ is associated with a corresponding pair
of rvs that are independent of pairs of rvs associated with
all the other pairs of terminals.  {\em All the rvs are assumed to take their
values in finite sets}. Following their observation of the
random sequences as above, the terminals in $\mathcal{M}$ are
allowed to communicate among themselves over a public noiseless
channel of unlimited capacity; all such communication, which
may be interactive and conducted in multiple rounds, is
observed by all the terminals. A communication from a terminal,
in general, can be any function of its observed sequence as
well as all previous public communication. The public
communication of all the terminals will be denoted collectively
by $\mathbf{F} = \mathbf{F}^{(n)}$.

The overall goal is to generate shared secret common randomness
for a given set $A \subseteq \mathcal{M}$ of terminals at the
largest rate possible, with the remaining terminals (if any)
cooperating in secrecy generation. The resulting secret key must
be shared by every terminal in $A$; but it need not be accessible
to the terminals not in $A$ and nor does it need to be concealed
from them. It must, of course, be kept secret from the
eavesdropper that has access to the public interterminal
communication $\mathbf{F}$, but is otherwise passive, i.e., unable
to tamper with this communication.\newpage

\vspace{0.2in}
\begin{figure}[h]
\begin{center}
\setlength{\unitlength}{1bp}%
\begin{picture}(200, 100)(0,0)
     \put(70,100){  $\widetilde{X}_2=(X_{2 1}, X_{2 3}, \ldots, X_{2 m})$ }
     \put(80,90){\circle*{5}}
     \put(90,85){2}
     \put(180,60){\circle*{5}}
     \put(0,70){  $\widetilde{X}_1 =(X_{1 2}, \ldots, X_{1 m})$  }
     \put(10,60){\circle*{5}}
     \put(15,50){1}
     \put(60,-20){ $\widetilde{X}_m =(X_{m 1}, X_{m 2}, \ldots, X_{m, m-1})$}
     \put(95, 80){\oval(200,70)}
     \put(200, 80){A}
     \put(175,30){\circle*{5}}
     \put(130,10){\circle*{5}}
     \put(70,0){\circle*{5}}
     \put(65,10){m}
\end{picture}
\end{center}
\end{figure}
\vspace{0.1in}
$\hspace{0.9in} \mbox{Figure 1: The PIN Model}$
\vspace{0.1in}


The following basic concepts and definitions are from \cite
{CsiNar04, CsiNar08}. Given $\epsilon >0$, for rvs $U, V$, we
say that $U$ is {\em $\epsilon$-recoverable} from $V$ if $Pr\{
U \neq f(V)\} \leq \epsilon$ for some function $f(V)$ of $V$.
With the rvs $K$ and $\mathbf{F}$ representing a secret key and
the eavesdropper's knowledge, respectively, information
theoretic secrecy entails that the security index\footnote{All
logarithms are to the base 2.}
\[ s(K;\mathbf{F}) = \log{|\mathcal{K}|} - H(K|\mathbf{F})\]
be required to be small, where $\mathcal{K}$ is the range of
$K$ and $|\centerdot|$ denotes cardinality. This requirement
simultaneously renders $K$ to be nearly uniformly distributed
and nearly independent of $\mathbf{F}$.

\textbf{ Definition 1:} Given any set $A \subseteq \mathcal{M}$
of size $|A| \geq 2,$ a rv $K$ constitutes an $\epsilon$-secret
key ($\epsilon$-SK) for the set of terminals $A$, achievable
with communication $\mathbf{F}$, if $K$ is
$\epsilon$-recoverable from $\left (\tilde{X}_i^n, \mathbf{F}
\right )$ for each $i \in A$ and, in addition, it satisfies the
secrecy condition
\begin{equation}
    s(K; \mathbf{F}) \leq \epsilon. \label{eqn:1}
\end{equation}

The condition (\ref{eqn:1}) corresponds to the concept of
``strong'' secrecy in which $\epsilon=\epsilon_n=o_n(1)$
\cite{Mau94, CsiNar04, CsiNar08}, as distinct from the earlier
``weak'' secrecy concept which requires only that $\epsilon_n =
o(n)$ \cite{Mau93, AhlCsi93}.

\textbf{ Definition 2:} A number $R$ is an achievable SK rate
for a set of terminals $A \subseteq \mathcal{M}$ if there exist
$\epsilon_n$-SKs $K^{(n)}$ for $A$, achievable with
communication $\mathbf{F}$, such that
\begin{equation}
    \epsilon_n \rightarrow 0\ \ \mbox{and}\ \
    \frac{1}{n}\log{|\mathcal{K}^{(n)}|} \rightarrow R\ \
    \mbox{as}\ \ n \rightarrow \infty.  \nonumber
\end{equation}
The largest achievable SK rate for $A$ is the SK capacity
$C(A)$.

Thus, by definition, the SK capacity for $A$ is the largest
rate of a rv that is recoverable at each terminal in $A$ from
the information available to it, and is nearly uniformly
distributed and effectively concealed from an eavesdropper with
access to the public interterminal communication; it need not
be concealed from the terminals in $A^c = \mathcal{M}
\backslash A$ that cooperate in secrecy generation.

A single-letter characterization of the SK capacity $C(A)$, $A
\subseteq \mathcal{M}$, for a general multiterminal source model,
of which the PIN model is a special case, is provided in
\cite{CsiNar04}. An upper bound for $C(A)$ in terms of
(Kullback-Leibler) divergence is also given therein and shown to
be tight in special cases.  These results play material roles
below.

\section{Results}

Our main results are the following. First, we obtain, upon
particularizing the results of \cite{CsiNar04}, a
(single-letter) expression for $C(A)$ for a PIN model,
in terms of a linear combination of mutual information terms
that involve only pairs of ``reciprocal'' rvs $\{ \left (X_{ij},
X_{ji}\right),\ 1 \leq i \neq j \leq m \}$. Second, stemming from
this observation, a connection is
drawn between SK generation for the PIN model and the combinatorial
problem of maximal packing of Steiner trees in an associated
multigraph.  Specifically, we show that the maximum rate of Steiner
tree packing in the multigraph is always a lower bound for SK
capacity.  Third, for the case $|A| = 2$ (when the Steiner
tree becomes a path connecting the two vertices in $A$) and for the
case $A=\mathcal{M}$ (when the Steiner tree becomes a spanning tree),
the previous lower
bound is shown to be tight.  This is done by means of an explicit
algorithm, based on maximal path packing and maximal
spanning tree packing, respectively, that forms an SK
out of independent SKs for pairs of
terminals.  In fact, the maximum rate of the SK thereby generated
equals the previously known upper bound for SK capacity
\cite{CsiNar04} mentioned above.

\subsection{SK Capacity}

We first give the SK capacity $C(A)$ for the PIN model. For
$A \subseteq \mathcal{M}$, let
\[\mathcal{B}(A) = \{ B \subset \mathcal{M}:\ B \neq \emptyset,\ B \nsupseteq A\} \]
and $\mathcal{B}_i(A)$ be its subset
consisting of those $B \in \mathcal{B}(A)$ that contain $i$,
$i \in \mathcal{M}$.
Let $\Lambda(A)$ be the set of all collections \\$\lambda =
\{\lambda_B:\ B \in \mathcal{B}(A)\}$ of weights $0 \leq
\lambda_B \leq 1$, satisfying
\begin{equation}
  \sum_{B \in \mathcal{B}_i(A)} \lambda_B = 1\ \ \
  \mathrm{for~all}\ \ \ i \in \mathcal{M}.  \label{eqn:2}
\end{equation} \newline

{\em \textbf{Proposition 3.1:} For a PIN model, the SK capacity
for a set of terminals $A \subseteq \mathcal{M}$, with $|A| \geq
2$, is

\begin{equation}\hspace{-2.8in} C(A) = \nonumber \end{equation}
\begin{equation}
    \hspace{0.2in} \min_{\lambda \in \Lambda(A)}
    \left [ \sum_{1 \leq i < j \leq m}
      \left (
        \mathop{\sum_{B \in \mathcal{B}(A):}}_{i \in B,\,j \in B^c}
        \lambda_B
      \right ) I(X_{ij} \wedge X_{ji})
    \right ].   \label{eqn:3}
\end{equation}}

{\em Remark:} (i) It is of interest in (\ref{eqn:3}) that
the SK capacity for a PIN model depends on the joint probability
distribution of the underlying rvs only through a linear combination
of the pairwise reciprocal mutual information terms.

(ii) We note from \cite[Theorem 3]{CsiNar04} that additional
independent randomization at the terminals in $\mathcal{M}$,
enabled by giving them access to the mutually independent rvs
$M_1, \ldots, M_m$, respectively, that are independent also of
$(\tilde{X}_1^n, \ldots, \tilde{X}_m^n)$, does not serve to
enhance SK capacity.  Heuristically speaking, the mentioned
independence of the randomization forces any additional ``common
randomness'' among the terminals in $A$ to be acquired only
through public communication, which is observed fully by the
eavesdropper.  On the other hand, randomization can serve to
enhance secrecy generation for certain models (cf. e.g.,
\cite{Wyner75})

{\bf Proof:} The proof entails an application of the formula
for SK capacity in \cite{CsiNar04, CsiNar08} to the PIN model.
For $B \in \mathcal{B}(A)$, denote $\tilde{X}_B = \left (
\tilde{X}_i,\ i \in B \right )$. From (\cite[Theorem
3.1]{CsiNar08},
\begin{equation}
  \hspace{-3.0in} C(A) = \nonumber
\end{equation}
\begin{equation}
  H\left ( \tilde{X}_{1}, \ldots,  \tilde{X}_{m}\right ) -
         \max_{\lambda\,\in\,\Lambda(A)}
           \sum_{B\,\in\,\mathcal{B}(A)}
             \lambda_B  H \left ( \tilde{X}_B | \tilde{X}_{B^c} \right). \label{eqn:4}
\end{equation} \newline
For the PIN model, since $\tilde{X}_i = \left ( X_{ij},\
j~\in~\mathcal{M} \backslash \{i\} \right),$ we observe in
(\ref{eqn:4}) that
\begin{eqnarray}
 H(\tilde{X}_1, \ldots, \tilde{X}_m)
    &=& H \left ( \{(X_{ij}, X_{ji})\}_{1 \leq i < j \leq m} \right ) \nonumber \\
    &=& \sum_{1 \leq i < j \leq m} H(X_{ij}, X_{ji}) \label{eqn:5}
\end{eqnarray}
and
\[\hspace{-1.0in} H(\tilde{X}_B | \tilde{X}_{B^c}) =
H(\tilde{X}_{\mathcal{M}}) - H(\tilde{X}_{B^c}) \]
\begin{eqnarray}
    &=& \sum_{1 \leq i < j \leq m} H(X_{ij}, X_{ji})
              - \mathop{\sum_{1 \leq i < j \leq m,}}_{i \in B^c,\,j \in B^c}
                H (X_{ij}, X_{ji}) \nonumber \\
    & &  - \sum_{i \in B^c,\,j \in B} H(X_{ij}) \nonumber \\
    &=& \mathop{\sum_{1 \leq i < j \leq m,}}_{i \in B,\,j \in B}
                H (X_{ij}, X_{ji}) +
        \sum_{i \in B,\,j \in B^c}
            H(X_{ij} | X_{ji}).  \label{eqn:6}
\end{eqnarray}
A straightforward manipulation of (\ref{eqn:4}),
using (\ref{eqn:5}), (\ref{eqn:6}), gives
\begin{eqnarray}
C(A) = \hspace{-0.2in} & & \min_{\lambda\,\in\,\Lambda(A)}~
    \sum_{1 \leq i < j \leq m} \Bigg [ H \left ( X_{ij}, X_{ji} \right ) \nonumber \\
& & \ \ \ \ \ \ \ \ \ \ \ \
   - \left ( \mathop{\sum_{B \in \mathcal{B}(A):}}_{i \in B,\,j \in B}
      \lambda_B \right ) H \left ( X_{ij}, X_{ji} \right ) \nonumber \\
& & \ \ \ \ \ \ \ \ \ \ \ \
   - \left ( \mathop{\sum_{B \in \mathcal{B}(A):}}_{i \in B,\,j \in B^c}
      \lambda_B \right ) H \left ( X_{ij} | X_{ji} \right ) \nonumber \\
& & \ \ \ \ \ \ \ \ \ \ \ \
   - \left ( \mathop{\sum_{B \in \mathcal{B}(A):}}_{i \in B^c,\,j \in B}
      \lambda_B \right ) H \left ( X_{ji} | X_{ij} \right ) \Bigg ]. \nonumber
\end{eqnarray}
Since, by (2),
\[\mathop{\sum_{B \in \mathcal{B}(A):}}_{i \in B,\,j \in B} \lambda_B
  = 1 - \mathop{\sum_{B \in \mathcal{B}(A):}}_{i \in B,\,j \in B^c} \lambda_B
  = 1 - \mathop{\sum_{B \in \mathcal{B}(A):}}_{i \in B^c,\,j \in B} \lambda_B,\]
we get
\begin{equation}\hspace{-2.8in} C(A) = \nonumber \end{equation}
\begin{equation}
    \hspace{0.2in} \min_{\lambda \in \Lambda(A)}
    \left [ \sum_{1 \leq i < j \leq m}
      \left (
        \mathop{\sum_{B \in \mathcal{B}(A):}}_{i \in B,\,j \in B^c}
        \lambda_B
      \right ) \left (\begin{array}{ll}
                    H(X_{ij} , X_{ji}) \\
                    - H(X_{ij} | X_{ji}) \\
                    - H(X_{ji} | X_{ij})
              \end{array} \right )
    \right ],   \nonumber
\end{equation}
thereby completing the proof.  $\qed$

\vspace{0.1in}
An upper bound had been established for SK capacity
for a general multiterminal source model \cite[Example
4]{CsiNar04}. This bound was expressed in terms of the
(Kullback-Leibler) divergence between the joint distribution of
the rvs defining the underlying correlated sources and the product
of the (marginal) distributions associated with appropriate
partitions of these rvs, thereby measuring the minimum mutual
dependence among the latter. The bound was particularized to the
PIN model in \cite{YeRez07}, and is restated below in a slightly
different form that will be used subsequently.

Let $\mathcal{P}$ be a partition of $\mathcal{M} =
\{1, \ldots, m\}$, and denote the number of atoms of $\mathcal{P}$ by
$| \mathcal{P} |$.\\

{\em \textbf{Lemma 3.2 \cite{YeRez07}:} The SK capacity $C(A),\ A
\subseteq \mathcal{M},$ for the PIN model is bounded above
according to

\[\hspace{-2.8in} C(A) \leq\]
\begin{equation}
  \hspace{0.1in} C^{ub}(A) \triangleq \min_{\mathcal{P}} \left ( \frac{1}{|\mathcal{P}|-1} \right )
    \left [
      \mathop{\sum_{1 \leq i < j \leq m}}_{(i,j)~\mbox{crosses}~\mathcal{P}} I(X_{ij} \wedge X_{ji})
    \right ],  \label{eqn:11}
\end{equation}
where for a fixed $\mathcal{P}$, a pair of indices $(i, j)$
crosses $\mathcal{P}$ if $i$ and $j$ are
in different atoms of $\mathcal{P}$.
The minimization in the right side of (\ref{eqn:11}) is over all
partitions $\mathcal{P}$ of $\mathcal{M}$ for which every atom
of $\mathcal{P}$ intersects $A$.}\\

\subsection{SK Capacity and Steiner Tree Packing}

There exists a natural connection between SK
generation for the PIN model and the combinatorial problem of
tree packing in an associated multigraph.

Let $G=\left (V, E\right )$ be a multigraph, i.e., a connected
undirected graph with no selfloops and with multiple
edges possible between any vertex pair, whose vertex set $V =
\mathcal{M} = \{1, \ldots, m\}$ and edge set
$E = \{e_{ij} \geq 0,~1 \leq i < j \leq m\}$, where
$e_{ij}$ is the number of edges
connecting the pair of vertices $i, j,~1 \leq i < j \leq m$.

\vspace{0.1in}

\textbf{Definition 3:} For $A \subseteq \mathcal{M}$, a {\em
Steiner tree} of $G$ (for $A$) is a subgraph of $G$ that is a tree
and whose vertex set contains $A$. A {\em Steiner packing} of $G$
is any collection of edge disjoint Steiner trees of $G$. Let $\mu(A,
G)$ denote the maximum size of such a packing (cf.
\cite{Die05}).

\vspace{0.1in}

We note that when $|A| = 2$, a Steiner tree for $A$ always
contains a path connecting the two vertices in $A$. Clearly, it
suffices to take $\mu(A, G)$ to be the
maximum number of edge disjoint paths connecting the two terminals
in $A$.

Next, assume without any loss of generality in the PIN
model that all pairwise reciprocal mutual information values
$I(X_{ij} \wedge X_{ji}),\ 1 \leq i \neq j \leq m,$ are
rational numbers.  Let $\mathcal{N}$
denote the collection of positive integers $n$ such that the
number of edges between any pair of vertices $i, j$ is equal to
$n I(X_{ij} \wedge X_{ji})$ is integer-valued for all $1 \leq i
\neq j \leq m$; clearly, the elements of $\mathcal{N}$
form an arithmetic progression. For a PIN model, consider a
sequence of associated multigraphs $\{G^{(n)} = \left
(\mathcal{M}, E^{(n)}\right ),\ n \in \mathcal{N}\}$, where
$E^{(n)},\ n \in \mathcal{N},$ is such that $e_{ij} = nI(X_{ij}
\wedge X_{ji})$.  We term $\sup_{n \in \mathcal{N}}
\frac{1}{n} \mu(A, G^{(n)})$ as the {\em maximum rate of Steiner tree
packing} in the multigraph $G = (\mathcal{M}, E)$.
The connection between SK generation for the
PIN model and Steiner tree packing is formalized below.\\

{\em \textbf{Theorem 3.3:} For a PIN model,

(i) the SK capacity satisfies
\begin{equation}
  C(A) \geq \sup_{n\,\in\,\mathcal{N}} \frac{1}{n}~\mu(A,
  G^{(n)}) \label{eqn:7}
\end{equation}
for every $A \subseteq \mathcal{M}$;

(ii) when $|A|=2$, the SK capacity is
\begin{eqnarray}
    C(A) &=& \sup_{n\,\in\,\mathcal{N}} \frac{1}{n}\
    \mu(A,
    G^{(n)}) \nonumber  \\
                   &=& C^{ub}(A). \label{eqn:14}
\end{eqnarray}}

{\em Remarks:} (i) The inequality in (\ref{eqn:7}) can be strict,
as shown by a specific example in a sequel paper \cite{Nitin_Nar}.
See also the remark following Theorem 3.4 for a heuristic explanation.

(ii) An exact determination of $\mu(A, G)$ is known to be NP-hard
\cite{Cheriyan_Salavatipour06}.  A nontrivial upper bound for
$\mu(A, G)$, similar in form to (\ref{eqn:11}), is known
\cite[paragraph 5 of Section 1]{Jian-etal03}. This bound can be
extended to yield an upper bound for $\sup_{n \in \mathcal{N}}
\frac{1}{n} \mu(A, G^{(n)})$ which, in general, is inferior to
that provided by $C(A)$ in (\ref{eqn:7}).


\vspace{0.1in}
{\bf Proof:}  (i) The proof consists of two main steps.  In the first
step, fix an $\epsilon >0$ that is smaller than every positive \newline
$I(X_{ij} \wedge X_{ji}),\ 1 \leq i < j \leq m$.
Each pair of terminals $i, j$ with $I(X_{ij} \wedge X_{ji}) > 0$,
generates a (pairwise) SK $K_{ij} = K_{ij}^{(n)}$ of size
$\lfloor n (I(X_{ij} \wedge X_{ji}) - \epsilon) \rfloor$ bits, using
public communication $F_{ij} = F_{ij}^{(n)}$, and satisfying
\begin{equation}
    s(K_{ij} ; F_{ij}) = o_n(1);   \label{eqn:10-a}
\end{equation}
the existence of such an SK follows from \cite{Mau94}.  The SK
achievability scheme in \cite{Mau94} consists of a ``weak'' SK
generated by Slepian-Wolf data compression, followed by ``privacy
amplification'' to extract a ``strong'' SK.  Note by the definition
of the PIN model that
$\{(K_{ij}, F_{ij})\}_{1 \leq i < j \leq m}$ are mutually
independent.

In the second step, consider the sequence of multigraphs \\
$\left \{ G^{(n)}_{\epsilon} =
(\mathcal{M}, \widetilde{E^{(n)}}) \right \}_{n=1}^{\infty}$, where
$\widetilde{E^{(n)}}$ is such that the number of
edges between any pair of vertices $i, j$ equals \\
$\lfloor n (I(X_{ij} \wedge X_{ji}) - \epsilon) \rfloor$.
We next show that every Steiner tree in a Steiner tree packing of
$G^{(n)}_{\epsilon}$ yields one shared bit for the terminals in $A$
that is independent of the communication in that Steiner tree.
Specifically, for edges $(i, j)$ and $(i, j'),\ j \neq j',$
with common vertex $i$ in the Steiner tree, vertex $i$
broadcasts to vertices $j, j'$ the binary sum of two
independent SK bits -- one with $j$ and the other with $j'$ --
obtained from the first step. This enables $i, j, j'$ to share
any one of these two bits, with the attribute that the shared bit is
independent of the binary sum. This method of propagation
(\cite[proof of Theorem 5]{CsiNar04}) enables all
the vertices in $A$, which are connected in the Steiner tree,
to share one bit that is independent of all the broadcast binary
sums from this tree.  Therefore, the maximum number of such shared
bits for the terminals in $A$ that can be generated by this
procedure equals $\mu(A, G^{(n)}_{\epsilon})$.
Denote these shared bits (of size $\mu(A, G^{(n)}_{\epsilon})$)
and the
communication messages generated by the mechanism in this
second step by $K = K^{(n)}(\{K_{ij}\}_{1 \leq i < j \leq m})$ and
$F = F^{(n)}(\{K_{ij}\}_{1 \leq i < j \leq m})$, respectively.

We claim that $K$ constitutes an SK for $A$.
Specifically, it remains to show that $K$ satisfies the secrecy condition
(\ref{eqn:1}) with respect to the overall communication in steps
1 and 2.  To this end, we denote
by $K_R^{(n)}(\{K_{ij}\}_{1 \leq i < j \leq m})$ all the pairwise
SK bits generated in the first step, that are residual from the maximal
Steiner tree packing of $G^{(n)}_{\epsilon}$ used to generate
$K$ by means of $F$.  Clearly,
\begin{equation}
    \{K_{ij}\}_{1 \leq i < j \leq m} = (K, F, K_R).  \label{eqn:9}
\end{equation}
Moreover, since the total number of edges in any Steiner tree
equals the sum of unity (i.e., the shared bit of $K$) and the number of bits
of public communication for that shared bit, we have
\begin{equation}
    |\widetilde{E^{(n)}}| = \log{|\mathcal{K}|} + \log{|\mathcal{F}|} + \log{|\mathcal{K}_R|},   \label{eqn:10}
\end{equation}
where $\mathcal{K}$, $\mathcal{F}$ and $\mathcal{K}_R$ denote
the respective ranges of $K$, $F$ and $K_R$.  Note that
$\log{|\mathcal{K}|} = \mu(A, G^{(n)}_{\epsilon})$.  Then,

\vspace{-0.1in}
\[\hspace{-1.7in} s(K;\{F_{ij}\}_{1 \leq i < j \leq m}, F)\]
\begin{eqnarray}
    &=& \log{|\mathcal{K}|} - H(K| \{F_{ij}\}_{1 \leq i < j \leq m}, F) \nonumber \\
    &\leq& \log{|\mathcal{K}|} - H(K| \{F_{ij}\}_{1 \leq i < j \leq m}, F, K_R) \nonumber \\
    &=& \log{|\mathcal{K}|}
        - H(K, F, K_R | \{F_{ij}\}_{1 \leq i < j \leq m}) \nonumber \\
    & &    + H(F, K_R | \{F_{ij}\}_{1 \leq i < j \leq m}) \nonumber \\
    &=& \log{|\mathcal{K}|}
        - H(\{K_{ij}\}_{1 \leq i < j \leq m} | \{F_{ij}\}_{1 \leq i < j \leq m}) \nonumber \\
    & & + H(F, K_R | \{F_{ij}\}_{1 \leq i < j \leq m}),\ \  \mbox{by~(\ref{eqn:9})} \nonumber
\end{eqnarray}
\begin{eqnarray}
    &\leq& \log{|\mathcal{K}|}
        + s(\{K_{ij}\}_{1 \leq i < j \leq m}; \{F_{ij}\}_{1 \leq i < j \leq m}) \nonumber \\
    & &    - |\widetilde{E^{(n)}}| + H(F, K_R) \nonumber \\
    &\leq& s(\{K_{ij}\}_{1 \leq i < j \leq m}; \{F_{ij}\}_{1 \leq i < j \leq m}), \ \  \mbox{by~(\ref{eqn:10})}  \nonumber \\
    &=& \sum_{1 \leq i < j \leq m} s(K_{ij}; F_{ij}), \nonumber \\
    &=& \frac{m(m-1)}{2} o_n(1), \nonumber
\end{eqnarray}
where the second-to-last equality is by the fact that
$\{(K_{ij}, F_{ij})\}_{1 \leq i < j \leq m}$ are mutually
independent, and the last equality is by (\ref{eqn:10-a}).
The maximum rate of the SK thus generated is equal to
$\lim_{n \rightarrow \infty} \frac{1}{n} \mu(A, G^{(n)}_{\epsilon})$
which, since $\epsilon >0$ was arbitrary, equals
$\sup_{n\,\in\,\mathcal{N}} \frac{1}{n}~\mu(A,
  G^{(n)}).$\\

(ii) Suppose that $A=\{1, 2\}$, and
note from the paragraph after Definition 3 that $\mu(A, G)$ is the
maximum number of edge disjoint paths in $G$ connecting terminals
$1$ and $2$.  It is clear that
$\frac{1}{n}\mu(A, G^{(n)})$ is nondecreasing in
$n \in \mathcal{N}$,
by the definition of $G^{(n)}$.  According to
Menger's theorem \cite{Menger, Bondy}, given a multigraph
$G=\left ( \mathcal{M}, E \right )$, the maximum number of
edge disjoint paths in $G$ connecting terminals $1$ and $2$
is equal to
\begin{equation}
  \mathop{\min_{\emptyset \neq B \subset \mathcal{M}}}_{1 \in B,\ 2 \in B^c}
                            \left (\mbox{number of edges that cross}~\{B, B^c\} \right ). \nonumber
\end{equation}
Applying this to $G^{(n)}$ as above, we have that for
$n~\in~\mathcal{N}$,

\begin{equation}
  \hspace{-2.5in}
  \frac{1}{n}\mu(A, G^{(n)}) =\nonumber
\end{equation}
\begin{equation}
\hspace{-0.0in} \frac{1}{n} \left [
                      \mathop{\min_{\emptyset \neq B \subset \mathcal{M}}}_{\ 1 \in B,\ 2 \in B^c}
                        \left ( \mathop{\sum_{1 \leq i < j \leq m:}}_{(i,j)~\mbox{crosses}~\{B, B^c\}}
                                     n I(X_{ij} \wedge X_{ji})
                        \right ) \right ]. \nonumber
\end{equation}
It then follows that
\begin{eqnarray}
  C(A) &\geq& \sup_{n\,\in\,\mathcal{N}}~ \frac{1}{n}\mu(A,
                        G^{(n)}),\ \ \ \ \mbox{by (8)} \nonumber \\
                &=& \mathop{\min_{\emptyset \neq B \subset \mathcal{M}}}_{1 \in B,\ 2 \in B^c}
                        \left ( \mathop{\sum_{1 \leq i < j \leq m:}}_{(i,j)~\mbox{crosses}~\{B, B^c\}}
                                     n I(X_{ij} \wedge X_{ji})
                        \right ) \nonumber \\
                &=& C^{ub}(A),  \ \ \ \ \mbox{by}~(\ref{eqn:11}).   \nonumber
\end{eqnarray}
The last equality follows upon noting that when $|A|=2$,
the minimization in (\ref{eqn:11}) is over only those partitions
that contain two atoms, each of which includes terminal 1 and terminal 2,
respectively.
This proves (ii). $\qed$

\subsection{SK Capacity and Spanning Tree Packing for $A = \mathcal{M}$}

When all the terminals in $\mathcal{M}$ seek a shared SK, i.e., when
$A = \mathcal{M}$, a Steiner tree for $A$ is a spanning tree for
$\mathcal{M}$.  In this case, we show that the lower bound for
SK capacity in Theorem 3.3 (i) is, in fact, tight.  Specifically, we
show that the algorithm in the proof of Theorem 3.3 yields an SK of
maximum rate that coincides with the upper bound for $C(\mathcal{M})$
in Lemma 3.2. \newline

{\em \textbf{Theorem 3.4:} For a PIN model, the SK capacity
$C(\mathcal{M})$ is
\begin{eqnarray}
    C(\mathcal{M}) &=& \sup_{n\,\in\,\mathcal{N}} \frac{1}{n}\
    \mu(\mathcal{M},
    G^{(n)}) \nonumber  \\
                   &=& C^{ub}(\mathcal{M}).  \label{eqn:12}
\end{eqnarray}}

{\em Remark:} When $A \subset \mathcal{M}$, Steiner tree packing
may not attain SK capacity.  In SK generation, a helper terminal in
$A^c$ helps link the user terminals in $A$ in complex ways through
various combinations of subsets of $A$.  In general, an optimal such
linkage need not be attained by Steiner tree packing.  However, when
$|A| = 2$, the two user terminals are either directly connected or are
connected by a path through helpers in $A^c$; both can be accomplished
by Steiner tree packing.  When $A = \mathcal{M}$, the mentioned
complexity of a helper is nonexistent.


\vspace{0.1in}
{\bf Proof:} The proof relies on a graph-theoretic result of
Nash-Williams \cite{Nas61} and Tutte \cite{Tut61}, that gives a
min max formula for the maximum size of spanning tree packing in a
multigraph.

It is clear that
$\frac{1}{n}\mu(\mathcal{M}, G^{(n)})$ is nondecreasing in
$n \in \mathcal{N}$,
by the definition of $G^{(n)}$. By \cite{Nas61, Tut61}, given a multigraph
$G=\left ( \mathcal{M}, E \right )$, the maximum number of
edge disjoint spanning trees that can be packed in $G$ is equal to
\begin{equation}
  \min_{\mathcal{P}} \Big \lfloor \frac{1}{|\mathcal{P}| - 1}
                            \left (\mbox{number of edges that cross }\mathcal{P} \right )
                     \Big \rfloor, \nonumber
\end{equation}
with the minimization being over all partitions $\mathcal{P}$
of $\mathcal{M}$.
Applying this to $G^{(n)}$ as above, we have that for
$n~\in~\mathcal{N}$,

\begin{equation}
  \hspace{-2.5in}
  \frac{1}{n}\mu(\mathcal{M}, G^{(n)}) =\nonumber
\end{equation}
\begin{equation}
\hspace{-0.10in} \frac{1}{n} \left [
                      \min_{\mathcal{P}} \
                         \Big \lfloor
                            \frac{1}{|\mathcal{P}| - 1}
                              \left (
                                \mathop{\sum_{1 \leq i < j \leq m:}}_{(i,j)~\mbox{crosses}~\mathcal{P}}
                                     n I(X_{ij} \wedge X_{ji})
                              \right )
                         \Big \rfloor
                    \right ]. \nonumber
\end{equation}
Denoting by $D$ the quantity in $\Big [  \ \Big ]$ above,
it follows that
\begin{eqnarray}
  C(\mathcal{M}) &\geq& \sup_{n\,\in\,\mathcal{N}}~ \frac{1}{n}\mu(\mathcal{M},
                        G^{(n)}),\ \ \ \ \mbox{by Theorem 3.3} \nonumber \\
                 &\geq& \sup_{n\,\in\,\mathcal{N}}~ \{D - \frac{1}{n}  \} \nonumber \\
                 &\geq& \min_{\mathcal{P}} \
                            \frac{1}{|\mathcal{P}| - 1 }
                              \left (
                                \mathop{\sum_{1 \leq i < i \leq m:}}_{(i,j)~\mbox{crosses}~\mathcal{P}}
                                     I(X_{ij} \wedge X_{ji})
                              \right ) \nonumber \\
                &=& C^{ub}(\mathcal{M}),
                                            \ \ \mbox{by}~(\ref{eqn:11}).
                                            \nonumber
\end{eqnarray}
The assertion in (\ref{eqn:12}) is now immediate. $\qed$

Lastly, the following observation is of independent interest.
Given a combinatorial problem of
finding the maximal packing of Steiner trees in a
multigraph, we can always associate with it a problem of SK
generation for an associated PIN model. By Theorem 3.3 (i),
the SK capacity for the
PIN model yields an upper bound for the maximum rate of
edge disjoint Steiner trees that can be packed in the multigraph;
the upper bound is tight both in the case of path packing by
Theorem 3.3 (ii) and in the case of
spanning tree packing by Theorem 3.4.

\section{Discussion}

Our proofs of Theorems 3.3 and 3.4 give rise to explicit
polynomial-time schemes for forming a group-wide SK for the
terminals in $A$ from the collection of optimum and mutually
independent SKs for pairs of terminals in $\mathcal{M}$ (namely
the $K_{ij}$s in the proof of Theorem 3.3). When $|A| = 2$ or $A =
\mathcal{M}$, our schemes achieve SK capacity.  Specifically, the
schemes combine known polynomial-time algorithms for finding a
maximal collection of edge-disjoint paths (resp. spanning trees)
connecting the vertices in $A$ when $|A| = 2$ (resp. $A =
\mathcal{M}$) \cite{Dinic, Edmonds_Karp, GarWes92} with the
technique for SK propagation in each tree as in the proof of
Theorem 3.3.

For a general multiterminal source model, the notions of wiretap
secret key (WSK) \cite{Mau90, AhlCsi93, CsiNar04} and private key
(PK) \cite{CsiNar04} have also been
proposed. Specifically, these notions involve an extra
``wiretapped'' terminal, say $m+1$, that observes $n$ i.i.d.
repetitions of a rv $\tilde{X}_{m+1}$ with a given joint pmf with
$(\tilde{X}_1, \ldots, \tilde{X}_m)$, and to which the eavesdropper
has access.  The key must now be concealed from the eavesdropper's
observations of $\tilde{X}_{m+1}^n = (\tilde{X}_{m+1, 1}, \ldots,
\tilde{X}_{m+1, n})$ and the public communication.  The notion of a
WSK requires that terminal $m+1$ not cooperate in key generation.
The less restrictive notion of a PK allows cooperation by terminal
$m+1$ by way of public communication.  The corresponding capacities
for the terminals in $A \subseteq \mathcal{M}$ are defined in the
usual manner, and denoted by $C_W(A)$ and $C_P(A)$.  We remark
that in the context of a PIN model, terminal $m+1$ represents a
compromised entity.

One model for the wiretapped rv $\tilde{X}_{m+1}$ entails its
consisting of $\left ( \begin{array}{cc} m \\ 2 \end{array} \right
)$ mutually independent components, one corresponding to each pair
$(X_{ij}, X_{ji}),\ 1 \leq i < j \leq m,$ of legitimate correlated
signals. This model is unresolved even in the simplest case of
$m=2$ terminals \cite{Mau93, AhlCsi93, CsiNar04, GohAnan07,
GohAnan08}. Instead, we consider a different model which depicts
the situation in which an erstwhile legitimate terminal $m+1$
becomes compromised. Specifically, the model now involves every
legitimate terminal $i$ in $\mathcal{M}$ observing $n$ i.i.d.
repetitions of the rv $(\tilde{X}_i, X_{i, m+1})$, while terminal
$m+1$ observes $n$ i.i.d. repetitions of $\tilde{X}_{m+1} = (
X_{m+1, j},\ j \in \mathcal{M} )$. We argue in the following
proposition that the WSK and PK capacities for this PIN model are
the same as the SK capacity of a reduced PIN model obtained by
disregarding terminal $m+1$ and with each legitimate terminal $i$
in $\mathcal{M}$ observing just $\tilde{X}_i^n$.

\vspace{0.1in} {\em \textbf{Proposition 4.1:}  It holds that
\begin{equation}
    C_W(A) = C_P(A) = C(A). \nonumber
\end{equation}}
\vspace{0.1in}

\textbf{Proof:}  We shall prove that
\begin{equation}
   \nonumber C(A) \stackrel{(a)}{\leq} C_W(A) \stackrel{(b)}{\leq} C_P(A) \stackrel{(c)}{\leq} C(A).
\end{equation}
The inequality $(b)$ is by definition.
Next, let $K = K(\tilde{X}_1^n, \ldots, \tilde{X}_m^n)$ be a
SK for $A$ achieved with communication
$\mathbf{F} = \mathbf{F}(\tilde{X}_1^n, \ldots, \tilde{X}_m^n)$
for the reduced PIN model.
Then $K$ is also a WSK since
\[\hspace{-1.8in} s \left (K; \mathbf{F}, ( X_{m+1, j}^n,\ j \in \mathcal{M} ) \right )\]
\begin{eqnarray}
    &=& \log{|K|} - H \left ( K | \mathbf{F}, ( X_{m+1, j}^n,\ j \in \mathcal{M} ) \right ) \nonumber \\
    &=& s(K; \mathbf{F}) + I(K \wedge ( X_{m+1, j}^n,\ j \in \mathcal{M} )| \mathbf{F}) \nonumber \\
    &=& o_n(1) \nonumber
\end{eqnarray}
since $I \left ( K, \mathbf{F} \wedge ( X_{m+1, j}^n,\ j \in \mathcal{M} ) \right ) = 0$, thereby
establishing (a).  In order to establish (c), we claim that every
achievable PK rate is an achievable SK rate for the reduced PIN model upon using
randomization at the terminals in $\mathcal{M}$; by remark (ii) after
Proposition 3.1, (c) then follows. Since $( X^n_{m+1, j},\ j \in \mathcal{M} )$
is independent of $(\tilde{X}_1^n, \ldots, \tilde{X}_m^n)$, any terminal in $\mathcal{M}$,
say terminal 1, can simulate $( X_{m+1, j}^n,\ j \in \mathcal{M} )$ and broadcast it to
all the terminals.  Next, each terminal $i$ in $\mathcal{M}$ can
simulate $X_{i, m+1}^n$ conditioned on
$( X_{m+1, j}^n,\ j \in \mathcal{M} ) = ( x_{m+1, j}^n,\ j \in \mathcal{M} )$.  This
second step of randomization is feasible since
$(\tilde{X}_1^n, \ldots, \tilde{X}_m^n), X_{1, m+1}^n, \ldots, X_{m, m+1}^n$
are conditionally mutually independent conditioned on
$( X_{m+1, j}^n,\ j \in \mathcal{M} ) = ( x_{m+1, j}^n,\ j \in \mathcal{M} )$.  Thus,
each terminal $i$ in $\mathcal{M}$ now has access to $(\tilde{X}_i^n,
X_{i, m+1}^n)$ while the eavesdropper observes $( X_{m+1, j}^n,\ j \in \mathcal{M} )$,
so that the reduced PIN model for SK generation can be used to simulate
a PIN model for PK generation with the given underlying joint pmf.
Thus, any achievable rate of a PK for $A$ in the {\em given} PIN model
for PK generation is an achievable rate of a PK for $A$ in the
{\em simulated} model.  Further, the latter PK is a
fortiori an SK for $A$ in the reduced PIN model with randomization
permitted at the terminals in $\mathcal{M}$.  This establishes (c).  $\qed$

\vspace{0.1in}
In the proof of achievability of SK capacity for the general
multiterminal source model in \cite{CsiNar04}, an SK of optimum
rate was extracted from ``omniscience,'' i.e., from a
reconstruction by the terminals in $A$ of {\em all} the signals $(
\tilde{X}_i^n,\ i \in \mathcal{M} )$ observed by the terminals in
$\mathcal{M}$. In contrast, the scheme in Theorem 3.3 (ii) (resp.
Theorem 3.4)  for achieving SK capacity for a PIN model with
$|A|=2$ (resp. $A=\mathcal{M}$) neither seeks nor attains
omniscience; however, we note that omniscience can be attained by
letting the terminals in $\mathcal{M}$ simply broadcast all the
residual bits left over from a maximal path packing (resp. maximal
spanning tree packing).

We close with the observation that in the proof of Theorem 3.3, the
SK bit generated by each Steiner tree in Step 2 is exactly
independent of the public communication in that tree.  Thus, if the
pairwise SKs in step 1 are ``perfect'' with zero security index,
then so is the overall SK for $A$. It transpires that for the PIN
model, there
is a tight connection between ``perfect secrecy generation'' and
``communication for perfect omniscience,'' redolent of the
asymptotic connection in \cite{CsiNar04}.

This new connection and the
role of Steiner tree packing in attaining perfect omniscience and
generating perfect secrecy are the subjects of a sequel paper \cite{Nitin_Nar}.

\vspace{0.1in}

\section{Acknowledgement}

The authors thank the anonymous referees for their helpful
comments.  P. Narayan thanks Samir Khuller for the helpful pointer
to \cite{GarWes92}.


\begin{thebibliography}{99}

\bibitem{AhlCsi93} R.~Ahlswede and I.~Csisz\'ar, ``Common
randomness in information theory and cryptography, Part I: Secret
sharing,'' {\it IEEE Trans. Inf. Theory}, vol.~39, pp.~1121-1132,
July 1993.

\bibitem{Bondy} A. Bondy and U. S. R. Murty, {\em Graph Theory,}
Series, Graduate Texts in Mathematics, Vol. 244: Springer, 2008.
\newpage

\bibitem{Cheriyan_Salavatipour06} J.~Cheriyan and M.~Salavatipour,
``Hardness and approximation results for packing Steiner trees.''
{\em Algorithmica}, vol. 45, pp. 21-43, 2006.

\bibitem{CsiNar04} I.~Csisz\'ar and P.~Narayan, ``Secrecy
capacities for multiple terminals," {\it IEEE Trans. Inf. Theory},
vol.~50, pp.~3047-3061, Dec.~2004.

\bibitem{CsiNar08} I. Csisz\'ar and P. Narayan, ``Secrecy
capacities for multiterminal channel models," {\it Special Issue
of the IEEE Trans. Inf. Theory on Information Theoretic Security,}
vol.~54, pp.~2437-2452, June 2008.


\bibitem{Dinic} E. A. Dinic, ``An algorithm for the solution of the
problem of maximal flow in a network with power estimation,'' {\em
Dokl. Akad. Nauk SSSR.} vol.~194, pp.~754-757, 1970.


\bibitem{Edmonds_Karp} J. Edmonds and R.M. Karp, ``Theoretical
improvements in algorithmic efficiency for network flow
problems,'' in {\em Combinatorial Structures and their
Applications}, New York: Gordon and Breach, 1970, pp.~93-96.

\bibitem{GarWes92} H. N. Gabow and H. H. Westermann: ``Forests,
frames, and games: algorithms for matroid sums and applications,''
{\em Algorithmica}, 7, pp.~465-497, 1992.

\bibitem{GohAnan07}
A. Gohari and V. Anantharam, ``Communication for omniscience by a
neutral observer and information-theoretic key agreement of
multiple terminals," in {\em Proc. 2007 IEEE Int. Symp. Inf.
Theory}, Nice, France, pp.~2056-2060.

\bibitem{GohAnan08}
A. Gohari and V. Anantharam, ``New bounds on the
information-theoretic key agreement of multiple terminals,'' in
{\em Proc. 2008 IEEE Int. Symp. Inf. Theory}, Toronto, Ontario,
Canada, pp.~742-746.

\bibitem{Die05} M. Gr\"otschel, A. Martin and R. Weismantel,
``Packing Steiner trees: A cutting plane algorithm and
computational results," {\em Math. Programming,} vol.~72,
pp.~125-145, Feb. 1996.

\bibitem{Jian-etal03} K. Jain, M. Mahdian, and M.R. Salavatipour,
``Packing Steiner trees,'' in {\em Proc. 14th ACM-SIAM Symp. on
Discrete Algorithms (SODA)}, Baltimore, Maryland, 2003,
pp.~266-274.

\bibitem{Mau90} U.~M. Maurer, ``Provably secure key
distribution based on independent channels,'' presented at the
{\em IEEE Workshop Inf. Theory}, Eindhoven, The Netherlands, 1990.

\bibitem{Mau93} U.~M. Maurer, ``Secret key agreement by public
discussion from common information,'' {\em IEEE Trans. Inf.
Theory}, vol.~39, pp.~733-742, May 1993.

\bibitem{Mau94} U.~M. Maurer, ``The strong secret key rate of
discrete random triples,'' in {\em Communications and
Cryptography: Two Sides of One Tapestry}, R.~E.~Blahut et al.,
Eds., Norwell, MA: Kluwer, Ch. 26, pp.~271-285, 1994.


\bibitem{Menger} K. Menger,
``Zur allgemeinen kurventheorie,'' {\it Fund. Math.}, vol.~10,
pp.~96-115, 1927.

\bibitem{Nitin_Nar} S. Nitinawarat and P. Narayan, ``Perfect
secrecy, perfect omniscience and Steiner tree packing,'' {\em IEEE
Trans. Inf. Theory}, to appear.

\bibitem{Nas61} C. St. J. A. Nash-Williams, ``Edge disjoint
spanning trees of finite graphs.'' {\it J. London Math. Soc.},
36, pp. 445-450, 1961.

\bibitem{RenWo03} R. Renner and S. Wolf, ``New bounds in
secret-key agreement: The gap between formation and secrecy
extraction,'' in {\em Proc. EUROCRYPT 2003, Lecture notes in
Computer Science, vol. 2656}: Springer-Verlag, 2003, pp. 562-577.

\bibitem{Tut61} W. T. Tutte, ``On the problem of decomposing a
graph into $n$ connected factors,'' {\it J. London Math. Soc.},
vol. 36, pp. 221-230, 1961.

\bibitem{Wyner75} A. D. Wyner, ``The wire-tap channel,'' {\it Bell
Sys. Tech. J.}, vol. 54, pp. 1355-1387, 1975.

\bibitem{YeRezSha06} C. Ye, A. Reznik and Y. Shah, ``Extracting
secrecy from jointly Gaussian random variables,'' in {\em Proc.
2006 IEEE Int. Symp. Inf. Theory}, Seattle, pp. 2593-2597.

\bibitem{YeRez07} C. Ye and A. Reznik, ``Group secret key
generation algorithms,'' in {\em Proc. 2007 IEEE Int. Symp. Inf.
Theory}, Nice, France, pp. 2896-2900.

\end{thebibliography}
\end{document}